\documentclass[a4paper]{article}
\usepackage{amsmath}
\usepackage{amssymb}

\newcommand{\lhaak}[1]{\left | #1\right |}

\newcommand{\half}{\frac{1}{2}}

\renewcommand{\arraystretch}{1.5}
\newcommand{\brk}[1]{\renewcommand{\arraystretch}{1}
\begin{tabular}{l}#1\\\end{tabular}\renewcommand{\arraystretch}{1.5}}

\begin{document}

\begin{titlepage}
\rightline{\large June 2003}
\vskip 2cm
\centerline{\Large \bf Have mirror micrometeorites been detected?}

\vskip 1.2cm
\centerline{\large R. Foot$^{a}$ and S. Mitra$^{b}$\footnote{
E-mail address: foot@physics.unimelb.edu.au, saibalm@science.uva.nl}
}

\vskip 0.7cm
\centerline{\large \it $^{a}$ School of Physics,}
\centerline{\large \it University of Melbourne,}
\centerline{\large \it Victoria 3010 Australia}
\vskip 0.5cm
\centerline{\large \it $^{b}$ Instituut voor Theoretische Fysica,}
\centerline{\large \it Universiteit van Amsterdam, 1018 XE Amsterdam,}
\centerline{\large \it The Netherlands}

\vskip 2cm
\noindent
Slow-moving ($v \sim 15$ km/s) `dark matter particles' have allegedly been 
discovered in a recent experiment. We explore the possibility that
these slow moving dark matter particles are small mirror matter 
dust particles originating from our solar system. Ways of further
testing our hypothesis, including the possibility of observing these
dust particles in cryogenic detectors such as NAUTILUS, are also discussed.

\end{titlepage}

Drobyshevski et al\cite{drob,drob2} have been searching for slow-moving
($v \sim 30\text{ km/s}$) dark matter
objects and have obtained some interesting positive results.
While they have interpreted their results in terms of Planckian
mass objects with electric charge $\sim 10 e$ (daemon hypothesis), we 
will suggest
an alternative interpretation in this note.

Their idea is very novel and straightforward. Dark matter could 
consist of massive
particles with interactions
strong enough to be captured by our solar system (or perhaps
be a component of the gas cloud from which our 
solar system was formed).
In this case such particles would have a velocity (relative to the Earth) of
order $30$ km/s. If its interactions are weak enough and/or the
dark matter particles are heavy enough then such particles will
not be stopped in the Earth's atmosphere and can enter the Earth's
surface with a velocity of order $30$ km/s. 
If such particles interact electromagnetically, then
they may 
appear as a sort of cosmic ray, yet potentially distinguishable
from ordinary cosmic rays because of their low velocity.
Thus, one needs to design a suitable detector capable of searching
for such slow-moving dark matter objects.

The detector devised by Drobyshevski et al
is very simple.  It
consists of a tinned iron box containing two transparent polystyrene
plates arranged horizontally one above the other separated by a 
distance of 7 cm.  Each plate is
coated on the 
downside with a layer of scintillator, ZnS (Ag) powder, 
with photomultiplier tubes at each end
(i.e. the top and bottom).
See Ref. \cite{drob,drob2} for more details. The passage of a charged
particle through each of these plates or the tin walls
can potentially be detected by the photomultiplier
tubes. The time difference between the pulses from
the top and bottom
photomultiplier tubes ($\Delta t$) will allow
a determination of the velocity of the particle through their
detector. A positive (negative) $\Delta t$ corresponds to 
the upper photomultiplier tubes
being triggered before (after) the lower photomulitplier tubes.
Background from cosmic rays and other conceivable backgrounds
should occur equally for positive and negative $\Delta t$ bins
(Drobyshevski et al use bins of duration $\Delta t = 20\mu s$).
This can be exploited by defining an up-down asymmetry,
$R_n$, ($n = 1, 2, 3 ...$) defined by
\begin{eqnarray}
R_n = \frac{ N(-20n \ \mu s < \Delta t < -20(n-1) \ \mu s )}{
N(20(n-1) \ \mu s  < \Delta t < 20n \ \mu s) }
\end{eqnarray}
Clearly in the absence of exotic slow moving particles one
expects $R_n$ to equal 1 (for each value of $n$). Any 
statistically significant 
deviation from unity
would be an interesting signal for such new particles, with
velocity of order $L/\Delta t$ (where $L$ is the size
of their tin box).
According to Ref.\cite{drob},
there is a statistically significant anomaly
occurring for $R_2$ (see figure 2 of Ref.\cite{drob}):
\begin{eqnarray}
R_2 = 0.3 \pm 0.15
\end{eqnarray}
However, the
total number of events was statistically small, 
roughly 70 events (in the $20 \ \mu s< |\Delta t| < 40 \ \mu s$ bins)
during a 700 hour exposure\cite{drob}, and is as yet unconfirmed
by any other experiment.
Some interesting hints for a seasonal variation were also
claimed\cite{drob,drob2}. 

Taken at face value these results represent intriguing  evidence for a
flux of slow moving ($v \sim L/\Delta t \sim 10-15\text{ km/s}$) 
long lived massive particles, with fairly large
penetrating ability and which interact sufficiently to produce an
observable burst of photons. 
One possible, more specific interpretation of 
the experimental
results is in terms of very heavy plankian mass objects with
charge $\sim 10 e$\cite{drob}. Another possibility is that this
experiment
has observed the impacts of mirror dust particles, i.e.
mirror micrometeorites, as we will shortly
explain.

Mirror matter is predicted to exist if nature exhibits
an exact unbroken mirror symmetry (for
reviews and more complete set of references, see Ref.\cite{review}). 
For each type of the ordinary
particle (electron, quark, photon etc) there is a mirror partner
(mirror electron, mirror quark, mirror photon etc), 
of the same mass. The two sets of particles form 
parallel sectors each with gauge symmetry $G$
(where $G = SU(3) \otimes SU(2) \otimes U(1)$ in the 
simplest case)
so that the full gauge group is $G \otimes G$.
The unbroken mirror symmetry maps
$x \to -x$ as well as ordinary particles into mirror
particles. Exact unbroken time reversal symmetry
also exists, with standard CPT identified as the product
of exact T and exact P\cite{flv}.

Ordinary and mirror particles can interact with each
other by gravity and via the photon-mirror
photon kinetic mixing interaction\footnote{Given the 
constraints of gauge invariance, renomalizability and mirror
symmetry it turns out\cite{flv} 
that the only allowed non-gravitational interactions
connecting the ordinary particles with the mirror particles
are via photon-mirror photons kinetic mixing, $\mathcal{L} = 
\frac{\epsilon}{2} F^{\mu \nu} F'_{\mu \nu}$, where $F^{\mu \nu}$
($F'_{\mu \nu}$) is the field strength tensor for electromagnetism
(mirror electromagnetism) and via a Higgs-mirror Higgs quartic
interaction, $\mathcal{L} = \lambda \phi^{\dagger} \phi \phi'^{\dagger}
\phi'$. If neutrinos have mass, then ordinary - mirror
neutrino oscillations may also occur\cite{flv2,f}.}, the effect of which is to
give mirror charged particles a small effective
ordinary electric charge $\epsilon e$\cite{flv,hol}.
Interestingly, the existence
of photon-mirror photon kinetic mixing
allows mirror matter to explain a number of puzzling
observations, including the pioneer spacecraft anomaly\cite{p1,p2},
anomalous meteorite events\cite{fy,doc}
and the unexpectedly low number of small craters on the
asteroid 433 Eros\cite{fm,eros}. 
It turns out that these 
explanations and other constraints\cite{ortho,fg2} suggest that
$\epsilon$ is in the range 
\begin{eqnarray}
10^{-9} \lesssim |\epsilon | \lesssim 10^{-6}.
\end{eqnarray}
In table \ref{tab} we have summarized the observational effects 
of the mirror world for $\epsilon$
in this range.

\begin{table}
\begin{scriptsize}
\begin{center}
\begin{tabular}{|l|l|l|}\hline
\brk{Observed phenomena/\\ prediction} & \brk{Observations consistent\\
with prediction?\\ Y/N}&
\brk{Preferred $\epsilon$ range}\\\hline
\brk{Dark matter}\cite{dark} & Y&-\\\hline

\brk{Microlensing by\\ mirror stars \cite{mstr}}& Y&-\\\hline

\brk{Mirror planets\\ orbiting stars \cite{mplnt}}& Y&-\\\hline

\brk{Ordinary planets\\ orbiting mirror stars \cite{isl}}& Y&-\\\hline

\brk{Orthopositronium-\\mirror orthopositronium\\ oscillations \cite{fg}} 
& ?\cite{ortho} &$\lhaak{\epsilon}\lesssim 
10^{-6}$\\\hline

\brk{Pioneer spacecraft\\ anomaly
\cite{pion, fm}}& Y \cite{p2}& $\lhaak{\epsilon}\gtrsim 10^{-9}$
\\\hline

\brk{Lack of small\\ craters on asteroids \cite{fm}}& Y \cite{eros}&
$10^{-6} \gtrsim \lhaak{\epsilon}\gtrsim 10^{-9}$
\\\hline

\brk{Anomalous meteoritic\\ events \cite{fy,fm}}& Y \cite{doc}&
$\lhaak{\epsilon}\gtrsim 10^{-9}$
\\\hline


\end{tabular}
\caption{Predicted effects of the mirror world.}\label{tab}
\end{center}
\end{scriptsize}
\end{table}

The properties of mirror matter space-bodies (SB) impacting on the
Earth's atmosphere has been studied in some detail in Ref.\cite{fm,fy}. 
Things depend (mainly) on the following parameters: the velocity
of the SB ($v$), the direction of its trajectory ($\cos \theta$),
the SB diameter ($D_{\text{SB}}$), and the value of the fundamental
parameter $\epsilon$.
(Of course, while the parameters $v, \cos\theta, D_{\text{SB}}$,
can all have different values, depending on each event, $\epsilon$
can only have one value which is fixed in nature, like the
fine structure constant).

While previous work\cite{fy,fm} focussed on the impacts of
fairly large objects ($\gtrsim 1$ meter)
which can potentially explain various anomalous impact
events (such as the Jordan event, tunguska event etc\cite{doc}), the possible
effects of small dust particle  
impacts on Earth was not
specifically explored. 
In fact, the number of small mirror dust particles could potentially be
quite large as collisions of large ($\gtrsim 1$ meter) mirror space
bodies with themselves and ordinary bodies will generate them
within our solar system.
Thus, such small particles are potentially important
because the number of such impacts on Earth should be much
greater than the impacts of larger bodies.
Furthermore, small dust particles can potentially
retain their cosmic velocity
impacting on the Earth's surface with
velocity of $11\text{ km/s}\lesssim v \lesssim 70\text{ km/s}$.
\footnote{The minimum impact velocity, 11 km/s, is equivalent to 
the escape velocity for
a particle on Earth, while the upper limit of about 70 km/s assumes that
the particle is bound within the solar system.}.

The condition that
small mirror dust particles pass through the atmosphere
without losing their velocity is that
(from Eq.22 of Ref.\cite{fm})\footnote{
Unfortunately Ref.\cite{fm} (and Ref.\cite{fy}) contains 
an error in the Rutherford cross section formula used ($\epsilon^2 e^4 \to
\epsilon^2 \alpha^2$). This means that the $\epsilon$ ranges
in Ref.\cite{fm} should increase by roughly one order of
magnitude.}

\begin{eqnarray}
|\epsilon | \lesssim 2 \times 10^{-7} \sqrt{\cos \theta} (v_i/30
\text{ km/s})^2
\end{eqnarray}
where $v_i$ is the initial velocity relative to the Earth.
Thus, since $v_i \lesssim 70\text{ km/s}$ (for solar system
objects), it follows that
such cosmic velocity impacts can begin to occur for
$|\epsilon | \lesssim 10^{-6}$.

At low velocities ($v \lesssim 70 \text{ km/s}$) the
energy loss is dominated by Rutherford scattering.
An important secondary process is bremsstrahlung,
where a real photon is emitted when the ordinary
nucleus (of electric charge $Ze$ and mass $M_A$) scatters 
off a mirror nucleus (of effective ordinary electric charge 
$\epsilon Z'e$ and mass $M_{A'}$).
The kinetic energy of a mirror iron nucleus (taking
the case of a mirror iron dust particle for definiteness)
moving at a velocity $v$ is:
\begin{eqnarray}
E_{\text{kin}} &=& \half M_{\text{Fe}} v^2
\nonumber \\
&\approx & 260 \left( \frac{v}{30\text{ km/s}}\right)^2 \ eV
\label{t2}
\end{eqnarray}
Clearly, there is sufficient energy to produce 
optical and UV photons via
the bremsstrahlung process.

In the case where 
\footnote{ Note that even in the case where 
$M_A \sim M_{A'}$, the cross section will
be of the same order of magnitude.}
$M_{A} \ll M_{A'}$, and for photon
energies ($k$) much less than $E_{\text{kin}}$,
the cross section 
is given by\cite{broj}
\footnote{
Unless stated otherwise, we use 
natural units, $\hbar = c = 1$.} 
\begin{eqnarray}
\left(\frac{d \sigma}{d\Omega} \right)_{\text{brem}} = \left( \frac{d\sigma}{
d\Omega}\right)_{\text{elastic}} \frac{2Z^2 \alpha}{\pi} 
\frac{4}{3} v^2 \sin^2 \frac{\theta_s}{2}
\ln \frac{k_{\text{max}}}{k_{\text{min}}}
\label{mon}
\end{eqnarray}
where the elastic cross section is given by\cite{merz}:
\begin{eqnarray}
\left(\frac{d\sigma}{d\Omega}\right)_{\text{elastic}} = 
\frac{4 M_A^2 \epsilon^2 \alpha^2 Z^2 Z'^2}{
(4M_A^2 v^2 \sin^2 \frac{\theta_s}{2} + \frac{1}{r_0^2})^2 }
\label{mon2}
\end{eqnarray}
where $r_0 \sim 10^{-9}$ cm is the radius at which the screening
effects of the atomic electrons becomes effective.
The total bremsstrahlung cross section can easily be
evaluated by integrating 
Eq.(\ref{mon})\footnote{
Note that the cross section, above, is valid assuming that the particles
are point particles.  
For our application, we have the scattering
of nuclei (of size $R_N$). In general we need to add in a Form factor,
but this effect is only important when the momentum
transfer is greater than $1/R_N \gtrsim 10$  MeV.
Given that we are concerned with solar system dust particles,
with relatively low velocity, $v \lesssim 70$ km/s,
the momentum transfer is always low enough so that we
can treat the atomic nuclei as point particles.},
\begin{eqnarray}
\sigma_{\text{brem}} = \frac{16\epsilon^2 \alpha^3 Z^4 Z'^2}{3 M_A^2 v^2}\ln (r_0 M_A v)
\ \ln \frac{k_{\text{max}}}{k_{\text{min}}}
\label{mon3}
\end{eqnarray}
Taking $k_{\text{min}} \sim $ 1 eV and
$k_{\text{max}} \sim 50$ eV, we find:
\begin{eqnarray}
\sigma_{\text{brem}} \approx
3 \times 10^{-32}
\left(\frac{\epsilon}{10^{-7}}\right)^2 \left( \frac{Z}{26} \right)^4
\left( \frac{Z'}{26} \right)^2 \left( \frac{50 M_P}{M_A} \right)^2
\left( \frac{10\text{ km/s}}{v}\right)^2 \text{ cm}^2
\label{last}
\end{eqnarray}
For the passage of a mirror iron nucleus through ordinary
matter of atomic number density, $n \sim 10^{23}/{\rm cm^3}$,
the mirror iron nucleus would have to
travel an average distance of $L = 1/(n \sigma) \sim 10^9$ cm 
before a bremsstrahlung process, producing a photon in
the energy range 1 eV $\lesssim k \lesssim k_{\text{max}}
\sim 50$ eV,
occurs.  However, a mirror dust particle contains, e.g.\ $ 10^{14}$
such mirror nuclei, so that $10^5$ photons (or more for 
larger dust particles) 
can  be produced when such a dust particle
passes through 1 cm of ordinary matter.

The importance of this is that these photons
can be detected in a photo-multiplier tube, and could
thereby explain the interesting experimental results
of Drobyshevski et al\cite{drob,drob2}. In this
interpretation, the excess of down-going events
occurs because of the
passage of a mirror matter dust particle impacting
with velocity $v \sim 10\text{ km/s}$.
The dust particle produces optical and UV photons via the bremsstrahlung
process upon passing through the detector. 
Hard elastic scattering in the scintillator,
may also produce an observable signal (depending on
their threshold, efficiency etc)
\footnote{
The seasonal dependence of
the flux, if real, might be due to the Earth passing through a mirror
matter dust stream, in much the same way that the Earth
passes through ordinary meteor streams. For example,
twice each year, in May and October we pass through the meteor stream
from Halley's comet. Perhaps there is a mirror meteor dust stream which we
pass through in August and February?}.


The flux estimated by Drobyshevski et al is $f \sim 10^{-5}
\text{ m}^{-2}\text{s}^{-1}$. This implies a solar system number density of 
order $n = f/v \sim
10^{-15}\text{ cm}^{-3}$ or one mirror dust particle per cubic kilometer
of the solar system, which seems plausible.
Such a tiny density of solar system mirror dust particles may
have been generated by random collisions of larger mirror space bodies.
Drobyshevski et al found an excess 
in the 10-15 km/s velocity region. This is close to the minimum
value expected, which might indicate a flux of particles moving 
in roughly circular orbits, near Earth's orbit. However
the bremsstrahlung cross section favours low velocities, since, 
from Eq.(\ref{last}),
$\sigma_{\text{brem}} \propto 1/v^2$.  
This feature, together with the relatively
low statistics collected so far, suggests that the distribution
of mirror dust particles might extend to higher velocites, potentially
up to the maximum ($v \approx 70$ km/s) for solar system particles.

The specific explanation presented here can 
be distinguished from the daemon hypothesis in a number of ways.
Perhaps the best way to do this would be to use an
up-down symmetric detector. 
The reason is that daemons are so heavy and compact that they can
penetrate the entire diameter of the Earth without losing
a significant proportion of their energy. This means
that the up-going daemon flux should be the same as
the down-going daemon flux. [In contrast, the mirror
dust particles stop in the Earth after a distance of 
$L \sim 10 (10^{-7}/\epsilon)^2 (v_i/30\text{ km/s})^4$ meters 
$\sim 10$ meters for $\epsilon \sim 10^{-7}$\cite{fm,f}].
Note that even though an up-down asymmetry was obtained in the
experiment, the detector was not up-down symmetric, which
could allow the daemon hypothesis to potentially explain the
results\cite{drob}. Further experimental
work, with an up-down symmetric detector should help distinguish
the daemon hypothesis from the mirror matter one.

The bremsstrahlung process, although important because it generates
easily detectable eV photons, is not the only way of detecting
mirror matter dust particles.  Most of the kinetic energy of
the dust particle is dissipated not via the bremsstrahlung process but
via Rutherford scattering. 
Let us now estimate the rate at which the kinetic
energy of a mirror dust particle is dissipated into
heat and vibration energy via Rutherford scattering. 
The rate of energy
loss is simply the product of the collision rate,  
the (forward) momentum lost per collision and the total
number of atoms in the mirror dust particle (${\cal N}$), which is
easily
evaluated to be 
\begin{eqnarray}
{dE_{\text{elastic}} \over dx} &=& -2{\cal N}\left( {M_{A'} \over M_A}
\right)
\int {d\sigma \over d\Omega} \rho v^2 \sin^2 {\theta_s \over 2} d\Omega
\nonumber \\
&\approx &
{4\pi {\cal N} Z^2 Z'^2 \rho \epsilon^2 \alpha^2 \over M_{A'} M_A v^2 } \ln (M_A v r_0)
\nonumber \\
& \sim &
\left({{\cal N} \over 10^{15}} \right)
\left( {\epsilon \over 10^{-7}}\right)^2 \left(
{30 {\rm km/s}\over v}\right)^2 \ 300 \ {\rm TeV/cm}
\label{gg}
\end{eqnarray}  
where $v$ (assumed to be $ \gtrsim 1$ km/s in the
above calculation) is the velocity of the
dust particle and $\rho$ is the mass density of ordinary matter
medium which the dust particle is moving through.

Eq.(\ref{gg}) can be compared with the energy loss due
to the bremsstrahlung process, 
\begin{eqnarray}
{dE_{\text{brem}} \over dx} &=& {\cal N} n \sigma_{\text{brem}} \langle k_{\gamma}
\rangle 
\nonumber \\ 
&\sim &\left({ {\cal N} \over 10^{15}} \right)
\left( {\epsilon \over 10^{-7}}\right)^2 \left( {30\ {\rm km/s} \over
v}\right)^2 \left( {\langle k_{\gamma} \rangle \over 10 \ eV}\right)
 \ {\rm MeV/cm}
\end{eqnarray}   
where $\langle k_{\gamma} \rangle$ is the mean energy of the 
bremsstrahlung photons emitted (of order 10 eV).
In fact, if one looks at the ratio, $(dE_{\text{brem}}/dx)/(dE_{\text{elastic}}/dx)$,
then the dependence on $\epsilon, {\cal N}, v$ cancels, leaving 
\begin{eqnarray}
{ dE_{\text{brem}}/dx  
\over 
dE_{\text{elastic}}/dx} &=& {4Z^2 \alpha M_{A'} \over 3\pi M_A^2} 
\ln {k_{\text{max}} \over k_{\text{min}}} 
\langle k_{\gamma} \rangle
\nonumber \\
& \sim & 10^{-9} 
\end{eqnarray}

The heat and vibration energy generated in
ordinary matter due to the passage
of mirror matter dust particles can potentially be observed 
in sensitive cryogenic detectors, such as the NAUTILUS gravitational wave
detector\cite{nautilus1}.  
NAUTILUS consists of an aluminum 2300 Kg bar cooled to 0.1 Kelvin.
In addition, there is a cosmic ray detector system. Although designed
to search for gravitational waves, NAUTILUS seems to be capable of
detecting mirror matter - type dark matter via detection of
the energy deposited in the bar via the elastic collisions
$(dE/dx)_{\text{elastic}}$, and potentially sensitive to $(dE/dx)_{\text{brem}}$
in the cosmic ray detector.
Interestingly, this
collaboration has found 
anomalously large energy depositing events in the bar which 
feature a paradoxically low electromagnetic component in
the cosmic ray detector\cite{nautilus}.
It seems to be possible that these anomalous events are associated
with the passage of a mirror dust particle, but it
may also be due to some unexpected property of aluminum\cite{nautilus2}.
Nevertheless it is interesting that the dark matter interpretation
of the St. Petersburg and NAUTILUS experiments
yield flux estimates which are roughly comparable:
\begin{eqnarray}
f_{\text{drob}} &\sim & 10^{-5} \text{ m}^{-2}\text{s}^{-1}
\nonumber \\
f_{\text{Nautilus}} &\sim & 2\times 10^{-6} \text{ m}^{-2}\text{s}^{-1}
\end{eqnarray}
The difference might be due to the detection thresholds being
different for the two experiments.

In conclusion, we have shown that mirror matter dust particles
impacting with the Earth
might appear as a sort of `anomalous cosmic ray'.
The anomalous features include a) it is slow-moving ($\sim 30$ km/s) and
b) the energy loss is dominated by elastic collisions. 
Interestingly, there is evidence for particles with these
anomalous features coming from two existing experiments.
Further work
and further experimental observations should clarify 
this interesting situation.

\vskip 1cm
\noindent
{\bf \large Acknowlegements:}
The authors would like to thank E. Drobyshevksi, for
helpful correspondence regarding his experiment and Z. Silagadze
for his comments on the paper.

\end{document}